\documentstyle[epsf]{article}
\oddsidemargin - 5mm
\begin{document}
\def\p{\phi}
\def\P{\Phi}
\def\a{\alpha}
\def\e{\varepsilon}
\def\be{\begin{equation}}
\def\ee{\end{equation}}
\def\l{\label}
\def\0{\setcounter{equation}{0}}
\def\b{\beta}
\def\S{\Sigma}
\def\C{\cite}
\def\r{\ref}
\def\ba{\begin{eqnarray}}
\def\ea{\end{eqnarray}}
\def\n{\nonumber}
\def\R{\rho}
\def\X{\Xi}
\def\x{\xi}
\def\la{\lambda}
\def\d{\delta}
\def\s{\sigma}
\def\f{\frac}
\def\D{\Delta}
\def\pa{\partial}
\def\Th{\Theta}
\def\o{\omega}
\def\O{\Omega}
\def\th{\theta}
\def\Ga{\Gamma}
\def\rar{\rightarrow}
\def\vp{\varphi}
\def\h{\hat}
\def\ga{\gamma}

\begin{flushleft}
{\bf (draft)}
\end{flushleft}
\begin{center}
{\Large\bf High-Multiplicity Processes}\\
({\it Theoretical status})
\vskip 0.5cm
{\bf J.Manjavidze\footnote{Permanent address: Inst. of Physics,
Tbilisi, Georgia}, A.Sissakian}\\ JINR, Dubna, Russia \end{center}

\begin{abstract}
We wish to demonstrate that investigation of
asymptotically high multiplicity (AHM) hadron reactions may solve, or
at least clear up, a number of problems unsolvable by other ways. We
would lean upon the idea: (i) the entropy is proportional to
multiplicity and, by this reason, at the AHM domain one may expect
the equilibrium final state and (ii) the high-multiplicity processes
becomes hard.  Last one means that the nonperturbative corrections
are frozen during this processes and the QCD predictions are
available for them.
\end{abstract}

\section{Introduction}\0

The interest to multiparticle processes noticeably falls down during
last decades and this trend is understandable: (a) the multiparticle
processes are too complicated because of large number of involved
degrees of freedom and (b) there is not quantitative hadrons theory
to describe the nonperturbative effect of color charge confinement.

It may be surprising at first glance that the asymptotically high
multiplicity (AHM) processes are simple at least from theoretical
point of view \C{sis1}. Nevertheless this is so and our aim is to
demonstrate this suggestion. We will consider also the simplest idea
of the trigger for AHM final state \C{sis2}.

We wish to note firstly that AHM states are equilibrium since the
corresponding entropy reach a maxim0um in the AHM region \C{landau}.
Then, a comparatively small number of simplest parameters is needed
for theirs description because of the Bogolyubov's correlations
relaxation principle.  Note also that in this case, following the
ergodic hypothesis, one may restricted by event-by-event
measurements, i.e. it is sufficient to observe small number of events
in the AHM domain. The cross sections in the AHM domain are assumed
extremely small, see e.g. \C{tevatron} and the last note seems
important from experimental point of view.

Besides, the AHM processes becomes hard and we may neglect (with
exponential accuracy) the background low-$p_t$ nonperturbative chanel
of hadrons production. Intuitively this is evident noting that in
all soft Regge-like theories, with interaction radii of hadrons
$\sim\sqrt{\ln s}$, one can not `stir' in the disk of area $\sim\ln
s$ arbitrary number of partons, since the high $p_t$ are cutted off.
Then the Regge-like theories are unable to describe the
multiplicities $n>\ln^2s$, if each parton in the disk is a source of
$\sim\ln s$ hadrons. At high energies the AHM domain is sufficiently
wide ($\ln^2s<<n_{max}\sim\sqrt s$) and, therefore, should be
`occupied' by hard processes \C{sis1}. Indeed, the mean transfer
momentum of created particle grows with $n_c$ \C{tevatron}.

One can say by this way that the confinement forces are frozen in the
AHM processes and one can expect that the pQCD predictions becomes
highly precise, forgetting on a moment the infrared (low-$x$) problem
of QCD.  We will return to this problem later.

So, the interest to AHM processes seems obvious, since (a) being
hard the experimental investigation of them may help to clear up the
structure of fundamental Lagrangian and (b) the theoretical
predictions in this field are more or less evident in the
asymptotically free theories. From our point of view the AHM
processes may be considered as a supplement to such popular physical
programs as the search of Higgs bosons, s-particles, $CP$ violation,
neutrino oscillations, the quark-gluon plasma (QGP) physics, the
early Universe dynamics.

We want to add also that in the AHM domain:\\
(i) the space density of colored constituents becomes large;\\
(ii) the created particles system becomes `cold' and the collective
effects becomes transparent;\\
(iii) the non-perturbative effects becomes essential only at the last
stage of the process.

It seems constructive to start investigation of hadron dynamics from
AHM domain and, experienced in this field, go to the moderate
multiplicities.  The to-day experimental status of AHM is following.
There is experimental data at TeVatron energies up to $n_c\approx5
\bar{n}_c (s)$ \C{tevatron}. The cross section $\s_{n_c}
\approx 10^{-5} \s_{tot}$ at this values of charged particles
multiplicity $n_c$ \C{tevatron} . The events with higher multiplicity
are unknown, if the odd cosmic data is not taken into account.  The
main experimental problem to observe AHM is smallness of
corresponding cross sections and impossibility to formulate
experimentally the `naive' trigger tagging the AHM events by charged
particles multiplicity. We think that this question is very important
since otherwise the AHM problem becomes pure academic.

We suggest to reject tagging the event by the multiplicity noting
that it is enough to have the `cold' final state in the AHM domain
\C{sis2}.  Then, to extract experimentally the AHM events the
calorimetric measurements are enough. Moreover, remembering that the
AHM states are equilibrium (or are near the equilibrium), the `rough'
measurements of small statistics should be enough. This practically
solves the problem of experimental investigation of AHM events.
For instance, such modern CERN experiments as ATLAS, CMS have
from this point of view good calorimeters. We will return to this
question later and bearing in mind such type `rough' experiments we
will follow corresponding ideology: formulating the theory we would
try to generalize ordinary inclusive approach \C{sis3}.

In Sec.2 we will give the physical interpretation of possible
asymptotics over $n$. In Sec.3 the hard Pomerons contribution in AHM
domain is discussed. In Sec.4 we will describe the QGP formation
signal at AHM. In Sec.5 the possible trigger for AHM is discussed.

\section{Classification of asymptotics over $n$}

It is better to start from pure theoretical problem assuming that the
incident total CM energy $E=\sqrt{s}$ is arbitrarily high and to
consider asymptotics over the total multiplicity $n$, assuming
nevertheless that $n<<n_{max}=\sqrt{s}/m$, where $m\simeq 0.2~Gev$
is the characteristic hadron mass, to exclude the influence of phase
space boundaries. Last one means that in the sum
$$
\X_{max}(z,s)=\sum_{n=1}^{n_{max}}z^n\s_n(s)
$$
we should choose the real positive $z$ so small that upper boundary
is not important and we can summing up to infinity:
\be
\X(z,s)=\sum_{n=1}^{\infty}z^n\s_n(s).
\l{1}\ee
This trick allows introduce classification of $\s_n$ asymptotics
counting the singularities of $\X(z,s)$ over $z$ \C{sis1}. So, if
$z_c$ is the solution of equation:
\be
n=z\f{\pa}{\pa z}\ln\X(z,s)
\l{2}\ee
then in the AHM region
\be
\s_n\sim e^{-n\ln z_c(n,s)}.
\l{3}\ee

To do further step we will use the connection with statistical
physics \C{elpat}. By definition
\be
\s_n(s)=\int d\o_n(q) \d(p_a+p_b-\sum_{i=1}^nq_i)|A_n|^2,
\l{4}\ee
where $A_n$ is the $a+b\rar(n~{\rm hadrons})$ transition amplitude
and $d\o_n(q)$ is the $n$ particles phase space element. There is
well known in the particle physics trick \C{satz} as this
$(3n)$-dimensional integrals may be calculated. For this purpose one
should use the Fourier transformation of the energy-momentum
conservation $\d$-function. Then, in the CM frame, if $n$ is large,
\be
\s_n(s)=\int \f{d\b}{2\pi}e^{\b E}\R_n(\b),
\l{5}\ee
where
\be
\R_n(\b)=\int d\o_n(q) \prod_{i=1}^ne^{-\b\e_i}|A_n|^2
\l{6}\ee
and $\e_i$ is the $i$-th particles energy.

Obviously this trick is used to avoid the constraints from the
energy-momentum conservation $\d$-function. But it have more deep
consequence. So, if we consider interacting particles $a$ and $b$ in
the black-body environment, then we should use the occupation number
$\bar{n}_{ext}$ instead of `Boltzmann factor' $e^{-\b\e_i}$.  For
bosons
\be
\bar{n}_{ext}(\b\e)=\{e^{\b\e}-1\}^{-1}.
\l{7}\ee
In result, replacing $e^{-\b\e_i}$ on $\bar{n}_{ext}$,
\be
\X(\b,z)=\sum_n z^n\R_n(\b)
\l{8}\ee
would coincide $identically$ with big partition function of
relativistic statistical physics, where $\b$ is the inverse
temperature $1/T$ and $z$ is the activity: the chemical potential
$\mu=T\ln z$ \C{sch}.

We may use this $S$-matrix interpretation of the $equilibrium$
statistical physics and consider $\X(z,s)$ as the energy
representation of the partition function.  Then, following to Lee and
Yang \C{lee}, $\X(z,s)$ should be regular function of $z$, for
practically arbitrary interaction potentials, in the interior of
unite circle. In the frame of natural physical assumption \C{lee,
kaz}, using estimation (\r{3}), one can find that
\be
\s_n(s)<O(1/n),
\l{9}\ee
i.e.  $\s_n(s)$ should decrease faster then any power of $1/n$.

First singularity may locate at $z=1$. It is evident from definition
(\r{1}) that $\X(z,s)$ would be singular at $z=1$ iff
\be
\s_n(s)>O(e^{-n}),
\l{10}\ee
i.e. in this case $\s_n(s)$ should decrease slower then any power of
$e^{-n}$. It follows from estimation (\r{3}), in this case $z_c(n,s)$
should be the decreasing function of $n$. Note that such solution,
at first glance, impossible. Indeed, by definition (\r{1}), $\X(z,s)$
should be increasing function of $z$ since all $\s_n$ are positive.
Then the solution of eq.(\r{2}) should be increasing function of $n$.
But, nevertheless, such possibility exist.

Remember for this purpose connection of $z$ with chemical
potential, $z_c(n,s)$ defines the work needed for creation of
additional particle. So, $z_c(n,s)$ may be the decreasing function of
$n$ iff the vacuum is unstable and the transition from false, free
from colorless particles (or being chiral invariant, etc.), vacuum to
the true one means colorless particles (or of the chiral-invariance
broken states, etc) creation.  Just this case corresponds to the
first order phase transition \C{lee, kaz}.

This phenomena describes expansion of the domain of new phase with
accelerating expansion of the new phase domains boundary, if the
radii of domain is larger then some critical value \C{callmen}. So,
$z_c(n,s)$ should decrease with $n$ since $z$ is conjugate to
physical particles number in the domain of new phase.

Following singularity in the equilibrium statistics may
locate at $z=\infty$ only. This is the general conclusion and means,
as follows from (\r{3}), that
\be
\s_n(s)<O(e^{-n}).
\l{11}\ee
In this case $\s_n(s)$ should fall down faster then any power of
$e^{-n}$. Note, investigation of the Regge-like theories gives just
this prediction \C{reggez}.

But considering the process of particles creation it is too hard
restrict ourselves by equilibrium statistics. The final state may be
equilibrium, as is expected in the AHM domain, but generally we
consider the process of incident (kinetic) energy dissipation into
particles mass. At very high energies having the AHM final state we
investigate the process of highly nonequilibrium states relaxation
into equilibrium one.

Such processes have interesting property readily seen in the
following well known model.  So, at the very beginning of this
century couple P.  and T.Euhrenfest had offered a model to visualize
Boltzmann's interpretation of irreversibility phenomena in statistics
\C{ehren}.  The model is extremely simple and fruitful.  It considers
two boxes with $2N$ numerated balls. Choosing number $l=1,2,...,2N$
$randomly$ one must take the ball with label $l$ from one box and put
it to another one.  Starting from the highly `nonequilibrium' state
with all balls in one box it is seen tendency to equalization of
balls number in the boxes.  So, there is seen irreversible flow
toward preferable (equilibrium) state. One can hope that this model
reflects a physical reality of nonequilibrium processes with initial
state very far from equilibrium. A theory of such processes with
(nonequilibrium) flow toward a state with maximal entropy should be
sufficiently simple to give definite theoretical predictions since
there is not statistical fluctuation of this flaw.

Following this model one can expect total dissipation of incident
energy into particle masses. In this case the mean multiplicity
$\bar{n}$ should be $\simeq n_{max}$ \C{landau}. But it is well known
that experimentally $\bar{n}<< n_{max}$. Explanation of this
phenomena is hidden in the constraints connected with the
conservation laws of Yang-Mills field theory. It is noticeable also
that this constraints are not so rigid as in integrable systems,
where there is not thermalization \C{zakh}. In the AHM domain we
expect the free from above constraints dynamics.  So, the AHM state
may be the result of dissipation process governed by `free' (from QCD
constraints) irreversible flow.

The best candidates for such processes are stationary Markovian ones.
They well described by so-called logistic equation \C{logi} and lead
to inverse binomial distribution with generating function
\be
\X(z,s)=\s_{tot}(s)\left\{\f{z_s(s)-1}{z_s(s)-z}\right\}^\ga,~\ga>0.
\l{a}\ee
The normalization condition $\pa\ln\X(z,s)/\pa z|_{z=1}=
\bar{n}_j(s)$ determines the singularity position:
\be
z_s(s)=1+\ga/\bar{n}_j(s).
\l{12}\ee
Note, $z_s(s)\rightarrow1$ at $s\rightarrow\infty$ since
$\bar{n}_j(s)$ should be increasing function of $s$.

Mostly probable values of $z$ tends to $z_s$ from below with rising
$n$:
\be
z_c(n,s)\simeq z_s-\ga/n=1+\ga(\f{1}{\bar{n}_j(s)}-\f{1}{n})
\l{b}\ee
This means that the vacuum of corresponding field theory should be
stable. It is evident that $\s_n$ decrease in this case as the
$O(e^{-n})$:
\be
\s_n\sim e^{-\ga n/\bar n_j}.
\l{13}\ee

The singular solutions of (\r{a}) type arise in the field theory, when
the $s$-chanel cascades (jets) are described \C{jets}. By definition
$\X(z,s)$ coincide with total cross section at $z=1$. Therefore,
nearness of $z_c$ to one defines the significance of corresponding
processes. It follows from (\r{b}) that both $s$ and $n$ should be
high enough to expect the jets creation. But the necessary condition
is closeness to one of $z_s$, i.e. the high energies are necessary,
and high $n$ simplifies only theirs creation. Note the importance of
jet creation processes in early Universe, when the energy density is
extremely high.

Expressing the `logistic grows law' the singular structure (\r{a})
leads to following interesting consequence \C{logi}. The energy
conservation law shifts the singularity to the right. For instance,
the singularity associated with two-jets creation is located at
$z_c^{(2)}(s)=z_c(s/4)>z_c(s)$.  Therefore, the multi-jet events will
be suppressed with exponential accuracy in the AHM domain if the
energy is high enough, since at `low' energies even the exponential
accuracy may be insufficient for such conclusion \C{sis1}. One may
assume that there should be the critical value of incident energy at
which this phenomena may realized. So the AHM are able to `reveal'
the jet structure iff the energies are high enough. In this sense the
AHM domain is equivalent of asymptotic energies.

Summarizing above estimations we may conclude that
\be
O(e^{-n})\leq\s_n< O(1/n),
\l{c}\ee
i.e. the soft Regge-like chanel of hadron creation is suppressed in
the AHM region in the high energy events with exponential accuracy.

\section{Hard Pomeron}

During last 50 years the contribution (Pomeron) which governs the
$s$-asymptotics of the total cross section $\s_{tot}(s)$ is stay
unsolved. The efforts in the pQCD frame shows that the $t$-channel
ladder diagrams from dressed gluons may be considered as the
dynamical model of the Pomeron \C{bfklz}.

The BFKL Pomeron is arise in result of summation, at least in the
LLA, of ladder gluon diagrams in which the virtuality of space-like
gluons rise to the middle of the ladder. To use the LLA this
virtualities should be high enough. Noting that the `cross-beams'
(time-like gluons) of the ladder are the sources of jets it is
natural that in the AHM domain the jet masses $q^2_i$ are large
enough and one can apply the LLA.

Consideration of Pomeron as the localized object allows conclude that
the multi-Pomeron contribution is $\sim1/k!$, if $k$ is the number of
Pomerons. This becomes evident noting that in the $t$ channel the
distribution over $k$ localized uncorrelated `particles' should be
Poisonian. This factorial damping should be taken into account in the
AHM domain.

Following to our above derived conclusion, number of `cross-beams'
(jets) should decrease with increasing $n$, and, therefore, in the
AHM domain the BFKL Pomeron, with exponential accuracy, should
degenerate into ladder with two `cross-beams' only. The virtuality of
time-like gluons becomes in this case $\sim\sqrt s$. So, the bare
gluons are involved at high energies at the AHM.  This solution is in
agreement with our general proposition that in the AHM domain the
particles creation process should be stationary Markovian.

We conclude that the AHM processes gave unique possibility to
understand as the BFKL Pomeron is builded up. But there is the
problem, connected with masslessness of gluons. So, the vertices of
time-like gluons emission are singular at the $q_i^2=0$. It is the
well known low-$x$ problem. In the BFKL Pomeron this singularity is
canceled by attendant diagrams of the `real' soft gluons emission.
However, this mechanism should destroyed when number of created
particles (i.e. of gluons) is fixed. Note, the solution of this
problem is unknown.

We hope to avoid this problem noting that the heavy jets creation is
dominate in the AHM domain. This idea reminds the way as the
infrared problem is solved in the QED (The emission of photons with
wave length much better then the dimension of measuring devise is
summed up to zero.) The quantitative realization of this possibility
for QCD is in progress now.

\section{QGP}

There is a question: can we modelling in the terrestrial conditions
the early Universe. The hot, dense, pure from colorless particles
quark-gluon plasma (QGP) is the best candidate for investigation of
this fundamental problem \C{qgpz}.

In our opinion \C{sis3} the plasma is a state of unbounded charges.
The `state' assumes presence of some parameters characterizes the
collective of charges and the `unbounded' assumes that the state is
not locally, in some scale, neutral (we discuss the globally neutral
plasma).

The ordinary QED plasma assumes that the mean energies of charged
particles is sufficiently high (higher then the energy of particles
acceptance), i.e. the QED plasma is `hot'. The QCD plasma in opposite
does not be `hot'(in corresponding energy scale) since the thermal
motion moves apart the color charges. This leads to sufficient
polarization and further `boiling' of vacuum, with creation of $q\bar
q$ pairs. So, the QCD plasma should be dense and at the same time
`cold' enough.

There is two principal possibilities to create such state. Mostly
popular is QGP plasma formation in the heavy ion-ion collisions at
high energies. It is believed that at the central (head-on)
collisions one cam observe the QGP in the CM central region of
rapidities. But there is not in to-day situation the unambiguous
(experimental) signals of QGP formation (number of the theoretical
possibilities are discussed in literature).

Other possibility opens the AHM region: since in this region the hard
channel of particles creation is favorable dynamically (at high
energies) one can try to consider the collective of color charges on
preconfinement stage as the plasma state. Because of energy-momentum
conservation this state would be `cold'. We should underline that
possible cold QGP (CQGP) formation is just the dynamical,
nonkinematical, effect:  the estimation (\r{c}) means that the
sufficient polarization of vacuum and its `boiling' effects are
insufficient, are frozen, at the AHM. So, considered CQGP remains
relativistic. This solves the problem of unbounded charges
formation\footnote{ To amplify this effect it seems resonable to
create AHM in the ion-ion collisions}.

But the question -- may we consider the collective of colored charges
created in the AHM events as the `state' -- remain opened. To-day
situation in theory is unable to give answer on this question (even
in the QCD frame, this will be discussed). But we can show the
experimentally controlled condition when same parameters may be used
to characterize this collective.

For instance, we may examine in what conditions the mean energy of
colored particles may be considered as such parameter. Let us return
to the definition (\r{5}) for this purpose. To calculate the integral
over $\b$ we will use the stationary phase method. Mostly probable
values of $\b$ are defined by equation of state:
\be
E=\pa \ln\R(\b,z)/\pa\b.
\l{14'}\ee
It is well known that this equation have positive real solution
$\b_c$. Then, as was noted above, $\b_c$ coincide with inverse
temperature $1/T$ and this definition of $T$ is obvious in the
microcanonical formalism of statistical physics.

The parameter $\b_c$ is `good', i.e. has a physical meaning, iff the
fluctuations near it are Gaussian (It should be underlined that the
value of fluctuations may be arbitrary, but the distribution should
be Gaussian).  This is so if, for instance,
\be
\f{\R^{(3)}}{\R}-3\f{\R^{(2)}\R^{(1)}}{\R}+2\f{(\R^{(1)})^3}{\R}
\approx0,
\l{14}\ee
where, for identical particles,
\ba
\R^{(k)}(\b_c,z)\equiv\f{\pa^k\R(\b_c,z)}{\pa\b_c^k}=
\n\\
(-z\f{\pa}{\pa z})^k\int
d\o_n(q)\prod^k_{i=1}\e(q_i)f_k(q_1,...,q_k;\b_c,z)
\l{15}\ea
and $f_k(q_1,...,q_k;\b_c,z=1)$ is the $k$-particle inclusive
cross section. Therefore, the (\r{14}) condition requires smallness
of energy correlation functions. It is the obvious in statistics
energy correlations relaxation condition near the equilibrium. One
can find easily the same condition for higher correlation functions.

This conditions establish the equilibrium, when knowledge of one
parameter ($\b_c$ in considered case) is enough for whole systems
description.  The analogous conditions would arise if other
parameters are considered. For instance, the (baryon, lepton, etc.)
charge correlations relaxation condition means the `chemical'
equilibrium.  The quantitative expression of this phenomena is
smallness of corresponding correlation functions.

This conditions are controllable experimentally. But it is hard to
expect that at finite values of $n$, where the cross sections are not
too small, above derived conditions are hold, even in the AHM region.
Later we will find more useable from experimental point of view
conditions making more accurate analyses.

\section{AHM events triggering}

Following to our main idea we would consider following solution of
the triggering problem \C{sis2}. Let $\e_i$ be the energy of $i$-th
particle (we did not distinguish particles), $i=1,2,...,n$, and let
us introduce $\e_{max}=max\{\e_i\}$.  Then, we have \be n\geq
n_{min}\equiv\f{E}{\e_{max}} \l{min}\ee if the equality \be \sum
\e_i=E, \l{enc}\ee where $E$ is the total incident energy, may be
established experimentally. So, choosing $\e_{max}<<E$ we examine the
high-multiplicity events. The approach assumes that it is
unimportant to know $n$ exactly in the high-multiplicity domain. We
will try to adjust the theory to such formulation of experiment.

Note, the trigger shrinks the phase space volume:
\be
d^n\o_{\e_{max}}=\prod^n_{i=1}\f{d^3q_i}{(2\pi)^3 2\e_i}\Th (\e_i
-\e_{max}),
\l{phs}\ee
So, the trigger depress the leading-particles creation first of all.
The effectiveness of this trigger was shown using the PYTHIA Monte
Carlo simulation.

Let $\e_\mu$ be now the energy measured in $\mu$-th calorimeter
cell (bean), $\mu=1,2,...,M$. Then, instead of (\r{enc}), we should
assume that
\be
\sum \e_\mu =E
\l{enc'}\ee
and, following to our idea, $\e_{max} =max\{\e_\mu\}$. Let us assume
also that $ \e_\mu=\sum^{n_\mu}_{i=1} \e_i,$ where $n_\mu$ is the
number of particles in $\mu$-th cell,
\be
\sum n_\mu=n.
\l{2a}\ee
Then
\be n\geq M \geq \f{E}{\e_{max}}\equiv n_{min}
\l{noe}\ee

Assuming particles identity,
\be
d^n\o_{\e_{max}}\sim\prod_\mu\left\{
\prod^{n_\mu}_{i=1}\f{d^3q_i}{(2\pi)^32\e_i}\Th (\sum^{n_\mu}_{i=1}
\e_i -\e_{max})\right\}.
\l{phe'}\ee
Note, $M\geq E/\e_{max}$.

The calorimeters can not overlap the whole range
of rapidities. If we assume that $E'$ is the $measured$ by
calorimeter energy ($E'\leq E$) than in the inequality (\r{noe}) we
should change $E\rar E'$. In this case we examine AHM in the fixed
domain of the rapidity, assuming that $E'$ is fixed.

Offered trigger constraints (i) the `peripheral' interactions and
(ii) the `short-range' fluctuation (in calorimeter cells dimension)
of particles densities. It was assumed that the number of calorimeter
cells $M>>n$. This is important in the AHM domain. Besides the
dimension of cells should be smaller then the cross section of the
QCD jets. Otherwise the condition $\e_\mu\leq\e_{max}$ may suppress
jets creation in the AHM domain, where, as was shown above, the jets
mass may be comparable with incident energy.

By this reason, having in mind the real calorimeters, the first-level
trigger should be weakened. Noting that the events have a tendency
become hard in the AHM domain one may ask to trigger the high
(comparable with $E$) $E_t$ events. The PYTHIA simulation shows that
the total number (including particles nonregistered in calorimeter)
grows with total energy $E_t$ registered in calorimeter.

Number of events (normalized on incident particles flow) with given
$\e_{max}$ is
\be
N(E, \e_{max})=\sum^{n_{max}}_{n=n_{min}} \s_n (E, \e_{max})
\l{ne}\ee
where $\s_n (E, \e_{max})$ is the cross section with constraints on
the particles energies. This constraint is fixed by $\Th$-functions
in (\r{phs}). The PYTHIA simulation shows that the distribution $\s_n
(E, \e_{max})$ have a maximum at $n=\bar{n}(E,\e_{max})$ rising with
$\e_{max}\rar0$ and $\bar{n}(E,\e_{max})>>n_{min}$ at LHC energies.

The topological cross section as was shown above is the important
quantity being sensible to structure of the QCD vacuum. On other
hand, its experimental measurement is the problematic task in the
AHM domain, even if $n_c$ only may be measured with definite accuracy.
In considered example,
\be
\s_n(s)=\int d\e_{max}\s_n (E, \e_{max}).
\l{tcs}\ee
Using this definition one can try to find $\s_n(s)$ from integral
quantity $N(E, \e_{max})$. For instance, differentiating over
$\e_{max}$ we find:
\ba
\f{\pa N(E, \e_{max})}{\pa\e_{max}}= n_{min}\s_{n_{min}}(E,\e_{max})+
\n\\
\sum^{n_{max}}_{n=n_{min}}\f{\pa\s_n (E, \e_{max})}{\pa\e_{max}}
\l{20}\ea
Let us assume now that just $n_{min}=n$ is the independent quantity.
Then $\e_{max}=\bar{\e}\equiv E/n$. With this definitions,
\be
\f{\pa N(E,\bar{\e})}{\pa\bar{\e}}= n\s_n(E,\bar{\e})+
\sum^{n_{max}}_{\nu=n}\f{\pa\s_\nu (E,\bar{\e})}{\pa\bar{\e}}
\l{21}\ee
remembering that
\be
\bar{\e}\equiv\f{E}{n}=\e_{max}.
\l{22}\ee
Just last equality presents problem: the maximal energy $\e_{max}$
may exceed the arithmetic average $E/n$ in the deep AHM domain only.
We are unable to examine this possibility quantitatively since the
PYTHIA generator gives too big uncertainty in this region of
multiplicities. But note that above described possibility is not
interesting since it is too hard to achieve the region, where
$\e_{max}={E}/{n}$, experimentally.

\section{Inclusive description}

The calorimetric measurement introduces averaging over particles
number, momentum, charges, etc. It is natural then to adjust the
theory to such type of experiment.

Having in mind the ion-ion collisions also let us consider the
$n$-into-$m$ particles transition. We find:
\ba
\R_{nm}(\b)=\f{(-1)^{n+m}}{n!m!}N_m(\b_i,\h\p)N^*_n(\b_f,\h\p)
\R_0 (\phi),
\l{2.9'}\ea
where $N_m(\b_i,\h\p)$ is the initial `temperature' $1/\b_i$
$n$-particles number operator:
\ba
N_m(\b,\h\p)= \int d\o_m (q)\prod^{m}_{k=1}d x_k d y_k
\n\\\times
e^{-\b\e(q_i)}e^{-iq_k (x_k-y_k)} \f{\d}{\d\phi_-(y_k)}
\f{\d}{\d\phi_+(x_k)}
\l{N_m}\ea
and
\be
\R_0 (\phi)=Z(\phi_+)Z^*(-\phi_-)
\l{2.10'}\ee
is the vacuum-into-vacuum transition probability in the environment
of external field $\p$.

Introducing new coordinates:
\ba
x_k=R_k+r_k/2,~~~y_k=R_k-r_k/2,~~~
\n\\
x'_k=R'_k-r'_k/2,~~~y'_k=R'_k+r'_k/2
\l{2.11'}\ea
we come naturally from (\ref{2.9'}) to definition of the Wigner
functions \C{carr}:
\ba
\R_{nm}(\b)=\f{1}{n!m!}\int d\o_m (q')d\o_n (q)
\n\\\times
\int \prod^{n}_{k=1}d R_k \prod^{m}_{k=1}d R'_k e^{-\e(q)(\b_i+\b_f)}
\n\\\times
W_{mn}(q,R;q',R'),
\l{2.12'}\ea
where
\ba
W_{nm}(q,R;q',R')=(-1)^{n+m}\prod^{n}_{k=1} N_{+}(q_k,R_k; \h\p)
\n\\\times
\prod^{m}_{k=1} N_{-}(q'_k,R'_k;\h \p)
\R_0 (\p).
\l{2.13'}\ea
The Wigner function has formal meaning in quantum case (it is not
positively definite), but in the classical limit it has the meaning of
ordinary in statistics phase-space distribution function. It obey the
Liouville equation conserving the phase space volume.

It is natural to introduce the generating functional weighing the
operator $N_{\pm}(R,q;\h\p)$ by the arbitrary `good' function $z(R,q)$:
\be
N_{\pm}(\a,z;\h\p)\equiv \int d R d\o_1 (q)e^{-\e\b}z(R,q)
N_{\pm}(R,q;\h\p).
\l{2.20'}\ee
In result, summation over all $n,m$ gives the generating functional
of Wigner functions in the temperature representation:
\ba
\R(\a,z)=e^{-N_+(\b_i,z_i;\h\p)-
N_-(\b_f,z_f;\h\p)}\R_0(\p)
\n\\
\equiv e^{- N(\b,z;\h\p)}\R_0(\p).
\l{2.23'}\ea
Introducing the black-body environment and summing over $n$ and $m$
the structure of generating functional (\r{2.23'}) is the same as in
the real-time finite-temperature field theories. This is the
simplest (minimal) choice of environment. One can consider another
organization of the environment, e.g. may consist from the correlated
particles as it happens in the heavy ion collisions (the nucleons in
ion may be considered as the quasi-free particles at high energies).

So, we construct the two-temperature theory (for initial and final
states separately). In such theory with two temperatures
the Kubo-Martin-Schwinger (KMS) periodic boundary
conditions applicability is not evident. Note, KMS boundary
condition play the crucial role in Gibbs thermodynamics since the
temperature in it is introduced just by this condition. In our
approach the KMS boundary condition arise as the consequence of
specially chosen environment (namely, of the black-body
environment).                   xx

The temperature was introduced as the parameter conjugate to
created particles energies. By this reason the uncertainty principle
restrictions should be taken into account. Considering the
Fourier-transformed probability $\R(\b,z)$ as the observable quantity
the phase-space boundaries are not fixed exactly, i.e. the 4-vector
$P$ can be defined with some accuracy only if $\b_i$ are fixed, and
vice versa. It is the ordinary quantum uncertainty condition. In the
particles physics namely the 4-vector $P$ is defined by experiment.
Let us find the condition when both $P$ and $\b$ may be the well
defined quantities, i.e. may be used for description of $\R(\a,z)$.
This is necessary if we want to use the temperature formalism in
particles physics also.

Note, in statistical physics such formulation of problem has no
meaning since the interaction with thermostat is assumed. In result
of this interaction the energy of system is not conserved, i.e. the
systems word line belong to the energy surface (it can be thin if the
interaction with thermostat is weak).

Considering the general problem of particles creation it is hard to
expect that the constant $\b_{i(f)}(E)$ is a `good' parameter, i.e. that
the factorization conditions (\r{14}) are hold. Nevertheless there
is a possibility to have the above factorization property in the
restricted space-time domains of size $L$. It is the so called
`kinetic' phase of the process when the memory of initial state was
disappeared, the `fast' fluctuations was averaged over and we can
consider the long- range `slow' fluctuations only.

In this `kinetic' phase one can use the `local equilibrium'
hypothesis in frame of which $\b_{i(f)}(E) \rar \b_{i(f)}(R,E)$,
where $R \in L_c$ and $L_c$ is the dimension of the $measurement$
(calorimeter) cell. Note, we always can divide the external particles
measuring device on cells since the free states are measured. Such
description of nonstationary media seems favorable in comparison
with traditional one.  It is natural to take
\be
L_c << L,
\l{con1}\ee
where $L$ is the characteristic thermal fluctuations dimension. It is
assumed that $\b_{i(f)}(R,E) =const.$ if $R \in L_c$. In the
equilibrium $L\rar\infty$.

By definition, $1/\b_{i(f)}(R,E)= \bar{\e}(R)$ is the mean energy of
particles in the cell with dimension $L_c$. The fluctuations in
$\bar{\e}(R)$ vicinity should be Gaussian.

The quantum uncertainty principle dictates also the condition:
\be
L_c >> L_q,
\l{con2}\ee
where $L_q$ is the characteristic scale of quantum fluctuations ($L_q
\sim 1/m$ for massive theories).

The `infrared unstable' situation means that
\be
L_q >> L.
\l{inf}\ee
One should underline that $L$ defines the scale of thermodynamics
fluctuations and, by this reason, the inequality (\r{inf}) points to
(unphysical) instability in the infrared domain.

So, if conditions (\r{con1}, \r{con2}) are hold we can use the Wigner
functions to describe the phase-space distributions, i.e. the
formalism has right classical limit in this case.

To introduce the scales $L,~L_c$ into formalism we can divide the $R$
4-space on the cells of $L_c$ dimension:
\be
\int dR=\sum_{r}\int_{L_c}dR,
\l{cel}\ee
where $r$ can be considered as the cells 4-coordinate. Assuming
that the inequality (\r{con1}) is hold we can assume that
$
\b=\b(r),~~z=z(q,r),
$
are the constants at least on the $L_c$ scale. With this definitions
\ba
\h N_{\pm}(\b,z;\p)\equiv \sum_r \int d\o_1 (q)e^{-\b(r)(\e(q)+\mu(r,q))}
\n\\\times
\int_{L_c} d R\h N_{\pm}(R,q;\p),
\l{Lc}\ea
where
$$
\mu(r,q)\equiv\f{1}{\b(r)}\ln z(r,q)
$$
is the local chemical potential.

We can use described above generalization of inclusive description to
investigate the tendency to equilibrium (total thermalization) in
kinetic phase considering the energy correlations between various
calorimeter cells.

On the more early pre-kinetic stages no thermodynamical
shortened description can be applied and the pure quantum description
(in terms of momenta only) should be used. For this purpose one
should expand $\R(\a,z)$ over operators $ N_\pm$ and the integrations
over $\a_i,~\a_f$ gives ordinary energy-momentum conservation
$\d$-functions, i.e.  defines the system on the infinitely thin
energy sheet.

\section{Conclusion}

To trigger the AHM events the restricted from above created particles
energies prescription was considered. Theoretically this means
introduction into the phase space differential measure products of
$\Th$-functions, see (\r{phs}). it is equivalent of introduction of
local activities
\be
z(q,r)=\Th(\e_{max}(r)-\e(q))
\l{c1}\ee
So, the choice of $z(q,r)$ introduces the model of given calorimeter
into the theory, i.e. allows to adjust the theory to given trigger.

It was consider the unsuccessful attempt to find topological cross
sections $\s_n$ varying $\e_{max}$, i.e. $z$. This idea is natural
noting definition:
\be
\s_n(s)=\f{1}{n!}\f{d^n}{dz^n}\X(z,s)|_{z=0},
\l{c2}\ee
see (\r{1}). So the attempts to find $\s_n$ varying activity have a
meaning. The realization of this idea crucially depends from concrete
choice of calorimeters.

But to investigate the phase transition phenomena we may use other
possibility. Noting that the system have a tendency to become
equilibrium in the AHM domain and noting that the Gibbs free energy
is $\sim\ln\X$, we can compare the heat capacity in the hadrons
and photons (or leptons) systems in the AHM domain. It is the direct
measurement of phase transition.

\vspace{0.2in}
{\Large \bf Acknowledgement}

The discussions of the inside of BFKL Pomeron with L.Lipatov and
E.Kuraev, and of the mini-jets creation in the semi-hard processes
with E.Levin was fruitful.  The modern experimental possibilities
explanation given by G.Chelkov, Z.Krumshtein, V.Nikitin was important
for us.  The PYTHIA simulation results gave us kindly M.Gostkin. We
are sincerely grateful to all of them. We would like to thank
V.G.Kadyshevski to the kind interest and helpful discussions.

\end{document}